\documentclass[journal]{IEEEtran} 
\usepackage{url}

\ifCLASSINFOpdf 
  \usepackage[pdftex]{graphicx} 
\else 
\fi 
\usepackage{amsmath,amssymb, amsthm} 
\usepackage{multirow} 
\usepackage{authblk} 
\usepackage{cite} 
\usepackage{subfig}
\usepackage{url} 
\usepackage{soul,color}
\usepackage[utf8]{inputenc}
\usepackage{pgf} 
\usepackage{arydshln}

\usepackage{color} 
\usepackage{footnote}
\renewcommand{\footnoterule}{%
  \kern -10pt
}
\usepackage{booktabs} 
\usepackage{todonotes} 
\usepackage{algorithm} 
\usepackage{algorithmic} 
\usepackage{amsmath}
\usepackage{bm}      
\usepackage{nccmath}
\usepackage{float}
\usepackage[font=small,skip=0pt]{caption}
\usepackage{blindtext}
\usepackage{multicol}
\usepackage{graphicx}
\usepackage{caption}
\makeatletter
\newcommand*{\rom}[1]{\expandafter\@slowromancap\romannumeral #1@}
\makeatother

\allowdisplaybreaks

\usepackage[algo2e,ruled,norelsize,lined,boxed,linesnumbered]{algorithm2e}

\newcommand{\comment}[1]{}

\newtheorem{theorem}{Theorem} 
\newtheorem{lemma}[theorem]{Lemma}

\definecolor{dukeblue}{rgb}{0.0, 0.0, 0.61}
\definecolor{mypink}{cmyk}{0, 0.7808, 0.4429, 0.1412}
\newcommand{\ff}{\color{black}}
\begin{document}

\title{
Predictive Multi-Microgrid Generation Maintenance: Formulation and Impact on Operations \& Resilience
} 

\author{Farnaz Fallahi\IEEEauthorrefmark{1}, Murat Yildirim\IEEEauthorrefmark{1}, Jeremy Lin\IEEEauthorrefmark{2}, Caisheng Wang\IEEEauthorrefmark{3}
	
	\thanks{\IEEEauthorrefmark{1} Industrial \& Systems Engineering, Wayne State University, Detroit, MI, USA.
	\IEEEauthorrefmark{2} Transmission Analytics, Phoenix, AZ.
	\IEEEauthorrefmark{3} Electrical \& Computer Engineering, Wayne State University, Detroit, MI, USA.
	farnaz.fallahi@wayne.edu, murat@wayne.edu, jeremylin@transmissionanalytics.net,
	cwang@wayne.edu.} 
\vspace{-10mm}
} 
\markboth{This work has been submitted to the IEEE for possible publication}%
{Shell \MakeLowercase{\textit{et al.}}: Bare Demo of IEEEtran.cls for Journals} 
\maketitle 
\begin{abstract}

Industrial sensor data provides significant insights into the failure risks of microgrid generation assets. In traditional applications, these sensor-driven risks are used to generate alerts that initiate maintenance actions without considering their impact on operational aspects. The focus of this paper is to propose a framework that i) builds a seamless integration between sensor data and operational \& maintenance drivers, and ii) demonstrates the value of this integration for improving multiple aspects of microgrid operations. The proposed framework offers an integrated stochastic optimization model that jointly optimizes operations and maintenance in a multi-microgrid setting. Maintenance decisions identify optimal crew routing, opportunistic maintenance, and repair schedules as a function of dynamically evolving sensor-driven predictions on asset life. Operational decisions identify commitment and generation from a fleet of distributed energy resources, storage, load management, as well as power transactions with {\ff the} main grid and neighboring microgrids. Operational uncertainty from renewable generation, demand, and market prices are explicitly modeled through scenarios in the optimization model. We use the structure of the model to develop a decomposition-based solution algorithm to ensure computational scalability. The proposed model provides significant improvements in reliability and enhances a range of operational outcomes, including costs,  renewables, generation availability, and resilience.
\end{abstract}

\begin{IEEEkeywords}
Condition-Based Maintenance, Microgrid Operations, Stochastic Programming, L-Shaped Decomposition
	\vspace{-4mm}
\end{IEEEkeywords}
{\ff
\section*{Nomenclature} 
\addcontentsline{toc}{section}{Nomenclature} 

\textbf{Sets:} 
\begin{IEEEdescription}[\IEEEusemathlabelsep\IEEEsetlabelwidth{$\mathcal{M}_ij$} ] 
\item [$\Omega$] Set of scenarios with $\omega \in \Omega$. 
\item [$\mathcal{T}$] Set of weeks within the planning horizon with $t \in \mathcal{T}$. 
\item [$\mathcal{H}$] Set of hours within a week with $h \in \mathcal{H}$.
\item [$\mathcal{M}$] Set of MGs with $m \in \mathcal{M}$.
\item [$\mathcal{G}$] Set of DERs with $i \in \mathcal{G}$.
\item [$\mathcal{G}^\mathrm{r},\mathcal{G}^\mathrm{nr}$] Set of renewable/non-renewable DERs.
\item [$\mathcal{G}^\mathrm{o},\mathcal{G}^\mathrm{f}$] Set of operational/failed DERs.
\item [$\mathcal{J}$] Set of DERs type (renewable/non-renewable) with $j \in \mathcal{J}$.
\item [$\mathcal{B}$] Set of batteries with $b \in \mathcal{B}$.
 \item [$N(.)$] Set of neighbouring MGs with active connection. 
\end{IEEEdescription} 

\textbf{Binary Decision Variables:} 
\begin{IEEEdescription}[\IEEEusemathlabelsep\IEEEsetlabelwidth{$soc^{\omega}\in$} ] 
 
\item[$\nu$] Preventive maintenance decision of operational DERs.  
\item[$z$] Corrective maintenance decision of failed DERs. 
\item[$x^{\mathrm{crew}}$ ] Maintenance crew visit decisions. 
\item[$x$] Commitment indicator of non-renewable DERs. 
\item[$\beta^{\mathrm{on}},\beta^{\mathrm{off}}$] Start up/shut down indicator of non-renewable DERs.
\item[$e^{\textbf{+}},e^{\textbf{-}}$]  Battery charging/discharging indicator. 
\item[$g^{\mathrm{p}},g^{\mathrm{s}}$] Purchasing/selling status of MGs from/to the grid. 
\item[$u^{\mathrm{p}},u^{\mathrm{s}}$] Purchasing/selling power status of MGs from/to other neighbouring MGs.
\end{IEEEdescription} 
\textbf{Continuous Decision Variables:} 
\begin{IEEEdescription}[\IEEEusemathlabelsep\IEEEsetlabelwidth{$soc^{\omega}_{m,b,t}$} ] 
\item[$\pi^{\textbf{+}},\pi^{\textbf{--}}$] Charged/discharged power in Battery. 
\item[$y^{\mathrm{gp}},y^{\mathrm{gs}}$] Purchased/sold power from/to the grid by MGs. 
\item[$y^{\mathrm{p}},y^{\mathrm{s}}$] Purchased/sold power between MGs.
\item[$y$] Power output of DERs. 
 \item[$\psi^{\mathrm{c}},\psi^{\mathrm{n}}$] Curtailed critical/non-critical load. 
\item[$soc$] Battery state of charge.
\end{IEEEdescription} 
\textbf{Parameters:} 

\begin{IEEEdescription}[\IEEEusemathlabelsep\IEEEsetlabelwidth{$B_{m,i,t,h}^ij$}] 
\item [$T$] Number of weeks within the planning horizon. 
\item [$H$] Number of hours within a week.  
\item [$M$] Number of MGs.  
\item [$G$] Number of DERs.  
\item [$B$] Number of batteries. 
\item [$G^\mathrm{r},G^\mathrm{nr}$] Number of renewable/non-renewable DERs within MMG.
\item [$G^\mathrm{o},G^\mathrm{f}$] Number of operational/failed DERs within MMG.
\item [$G^\mathrm{r}_m,G^\mathrm{nr}_m$] Number of renewable/non-renewable DERs of MG $m$.
\item [$G^\mathrm{o}_m,G^\mathrm{f}_m$] Number of operational/failed DERs of MG $m$.
\item [$D^{\mathrm{c}},D^{\mathrm{n}}$] Critical/non-critical load. 
\item [$Y^{\mathrm{o}},Y^{\mathrm{f}}$] Preventive/corrective maintenance duration. 
\item [$\lambda^{\mathrm{c}},\lambda^{\mathrm{n}}$] Per unit cost of curtailed critical/non-critical load. 
\item[$\overline{P^\mathrm{{soc}}},\underline{P^\mathrm{{soc}}}$] Maximum/Minimum state of charge of a battery. 
\item [$\overline{P^\mathrm{{ch}}},\overline{P^\mathrm{{dch}}}$] Maximum charging/discharging rate of a battery.
\item [$\eta$] Batteries charging efficiency. 
\item [$C^\mathrm{{crew}}$] Maintenance crew deployment cost.
\item[$C$] Dynamic maintenance cost of DER. 
\item [$V$] No-load cost of non-renewable DERs . 
\item[$C^{\mathrm{on}},C^{\mathrm{off}}$] Start up/shut-down cost of conventional DERs. 
\item[$MU,MD$] Minimum up/down time of conventional DERs. 
\item[$RU,RD$] Ramp-up/down rate of conventional DERs. 
\item [$\overline{P^{\mathrm{nr}}},\underline{P^{\mathrm{nr}}}$] Maximum/minimum production capacity of conventional DERs.
\item [$\overline{F^{\mathrm{gp}}},\overline{F^{gs}}$] Maximum power a MG can purchases from/sells to the grid.
\item [$\overline{F^{\mathrm{p}}},\overline{F^{s}}$] Maximum power a MG can purchases from/sells to other neighbouring MGs.
\item [$\Phi$] Available production capacity of renewable DERs.
 \item [$\Gamma^{\mathrm{gp}},\Gamma^{\mathrm{gs}}$] Purchasing/selling electricity price from/to the grid.  
\end{IEEEdescription} }
\section{Introduction} 

Aging infrastructure, operational uncertainty, and the increasing requirements on reliability and
resilience have unleashed a flood of interest in the concept of microgrids.
Through enabling a focused control of distributed energy resources (DERs), microgrids effectively incorporate renewable resources, conventional generators (CGs), energy storage devices (SDs), and flexible local loads to serve the local network, and contribute to the grid\cite{hatziargyriou2014microgrids}.
Management of DERs in a microgrid, or a multi-microgrid (MMG) setting comes with its own set of unique challenges. Microgrids are responsible for the control of a heterogeneous fleet of DERs, the satisfaction of various local requirements, and continuous interaction with the grid and other microgrids; while being subjected to a high level of uncertainty from demand, renewable generation, and market prices. At the presence of high operational complexity, microgrid maintenance management has an increasing impact on microgrid operations. The critical impact of maintenance decisions motivates the use of sensor-driven condition monitoring approaches that provide additional visibility on generation outage risks. These generation outage risks are often used to trigger imminent maintenance actions on high-risk DERs without considering operational outcomes. An alternative approach based on proactive maintenance would identify early signs of degradation to conduct integrated operations and maintenance optimization ahead of time. To date, this integration proved challenging to implement in a microgrid setting.

Efficient management of microgrid operations is a fundamental challenge that revolves around optimizing power generation, energy trade, and storage management to supply forecasted local load and enhance resilience \& reliability. Microgrid operation modeling literature is rich, covering renewable penetration \cite{wang2018two,zhang2013robust}, operational uncertainties \cite{liu2017microgrid}, resilience \cite{khodaei2014resiliency,gholami2016microgrid}, and market interactions\cite{parhizi2016market}. In an MMG setting, the focus shifts to modeling the interactions between microgrids either in collaborative \cite{zhang2018robust,rahmani2018cooperative} or competitive \cite{asimakopoulou2013leader,ma2016multi} environments. Maintenance is often modeled by introducing additional constraints over the operations models. Typically, these constraints enforce periodic maintenance requirements over DERs \cite{mazidi2017strategic,ding2012opportunistic}. A key focus in the literature is to provide improvements to operations and maintenance by primarily focusing on operational aspects. In the presence of strong coupling between operations and maintenance, potential improvements over maintenance can play an equally crucial role in improving operational outcomes.

An important direction for improving maintenance policies is to leverage  sensor data. Sensor-driven policies offer significant advantages over conventional models by providing better predictions on asset failure risks. These risk predictions rely on condition monitoring techniques that monitor indicators of asset degradation from raw sensor inputs, such as vibration, temperature, and performance \cite{jordan2017photovoltaic,carden2004vibration}. Condition monitoring systems are actively used in wind turbines (WT), photovoltaics (PV), and conventional generators \cite{mahani2018joint,byon2010optimal,yildirim2016sensor,yildirim2017integrated}. For a comprehensive review of condition monitoring methods in power generation, readers are referred to \cite{hameed2009condition}. In traditional applications, condition monitoring systems alert the operators when the failure risks of DERs reach a certain severity. Depending on the subjective judgment of the maintenance personnel, these alerts are often used to initiate immediate maintenance actions. Such \emph{ diagnostic-based} policies rely solely on the current state of DERs and do not provide advanced notice for planning maintenance actions, which may pose significant risks to operations. 

\emph{Prognostic-based} policies provide the capability to derive dynamic predictions on remaining life distribution (RLD) throughout the asset lifetime, 
generating advanced notice for failure risks to enable a more proactive set of maintenance policies. Majority of prognostic-based approaches focus on single asset systems\cite{byon2010optimal}. Markov chain models are used to characterize degradation and derive optimal maintenance decisions \cite{mahani2018joint}. This literature is extended by more detailed approaches that jointly consider maintenance and environmental factors \cite{byon2010season}. While benefiting from prognostic predictions, these approaches do not necessarily consider the interactions across multiple generation assets and power system operations. Recently, \cite{yildirim2016sensor,Yildirim2016SensorDrivenCG,yildirim2017integrated} proposed joint operations and maintenance scheduling models for transmission networks. However, these models
do not consider any operational and market uncertainty that is vital to microgrid operations. It is an open problem to investigate the impact of prognostic-based maintenance policies on revenue, reliability, and resilience in complex and highly stochastic operational environments.

{In this paper, we propose a sensor-driven integrated framework that incorporates} i) real-time degradation models for a heterogeneous fleet of DERs, with ii) stochastic operations and maintenance models for large scale MMG systems. 
Sensor-driven degradation models continuously update the RLD predictions to identify asset-specific optimal maintenance decisions. The proposed sensor-driven integrated operations and maintenance scheduling model (SD-IOM) uses these predictions to derive fleet-optimal maintenance actions that minimize the MMG system operations and maintenance cost. Operational decisions in the SD-IOM determine unit commitment, generation dispatch, power transactions across microgrids and with the grid, storage scheduling as well as the load management decisions. {The joint modeling of detailed operations} with maintenance scheduling enables explicit characterization for the impact of sensor-driven maintenance on a range of operational outcomes, such as renewable integration, storage management, MMG reliability and resilience, and MMG contribution to the generation availability of the grid. Unique aspects of our methodology can be listed as follows:
\begin{itemize}
    \item We propose a sensor-driven framework that fuses degradation analytics with a stochastic optimization model for operations and maintenance in  MMG systems. Unique to our framework, is the integration of stochastic degradation models for asset remaining life prediction, with MMG operations and maintenance decisions, that include decisions within microgrid such as storage, generation, and load management; and decisions across microgrids such as power transactions, and maintenance crew visits.  
    \item {We use sensor data to derive generation failure risks and the associated dynamic maintenance cost functions; and build uncertainty scenarios for a wide range of maintenance and operational outcomes: i.e. renewable generation, demand, market prices, and connectivity stages to investigate the relative importance of these factors in terms of revenue, generation availability, grid contribution, and resilience.}
    \item We build a two-stage reformulation of the optimization model that decomposes the problem across maintenance periods and uncertainty scenarios. The proposed reformulation drastically reduces scalability problems due to the number of scenarios. We leverage on this structural property to devise a solution methodology that iteratively approaches the optimal solution through L-shaped and integer cuts. 
\end{itemize}

We develop a comprehensive microgrid operations and maintenance platform that uses real-world vibration-based degradation signals, PJM market prices, and operational data from NREL and NOAA. We conduct extensive sets of experiments to highlight the performance of our model in different settings in terms of reliability, maintenance performance, and operations. As a case in point, we show that the proposed maintenance policy can be as useful as additional storage capacity in terms of enhancing the resilience of microgrids.

The rest of the paper proceeds as follows. In section \rom{2}, we introduce our methodology that integrates a sensor-driven degradation modeling approach for DERs, with a stochastic mixed-integer optimization model for microgrid operations and maintenance. Section \rom{3} develops a reformulation and a solution methodology to enhance computational scalability. An extensive set of experiments are conducted in Section \rom{4} to showcase the performance of the proposed model. Finally, section \rom{5} provides conclusions and closing remarks.

\vspace{-4mm}
\section{Methodology} 
In this section, we introduce a unified framework for sensor-driven generation maintenance scheduling in multi-microgrid systems. The framework is composed of two subsections: sensor-driven degradation models that derive dynamic predictions on RLD and maintenance costs associated with each DER, and an adaptive optimization for opportunistic maintenance and operations scheduling. 
\vspace{-4mm}
 \subsection{Sensor-Driven Degradation Analytics}\label{section:DegradationAnalytics}
Generation asset performance and health deteriorate over time due to aging and wear - a process called degradation. 
Sensor readings can be utilized to discover implicit manifestations of this deterioration over many different energy assets, including WTs, PV panels, and CGs. Through measuring parameters such as vibration, temperature, light, etc, one can identify \textit{degradation signal} of a DER, a quantity that describes the real condition of the unit and is the basis to predict the future trajectory of degradation
\cite{gebraeel2005residual}. {\ff Over the DER's life, the degradation severity increases until it exceeds a predetermined failure threshold, which corresponds to the DER's failure. Although various DERs exhibit different degradation rates and failure times, those of the same type typically have a common degradation signal form. We use a continuous-time continuous-state parametric stochastic function to model the evolution of DERs degradation signals over time.
Degradation signal of DER  $i$ of type $j$ at week $t$, $D^{j}_{m,i}(t)$, is modeled as follows: 
\begin{equation}
 D^{j}_{m,i}(t)=h^j_{i}(\kappa^j,\phi_{m,i}^j,t)+\epsilon(t)
\end{equation}
 where $h^j_{i}(.)$ and $\epsilon(.)$ denote the general degradation function and the error term, respectively. 
 The parameters $\kappa^j$ and $\phi_{m,i}^j$  are the deterministic and stochastic degradation parameters, respectively. The deterministic parameter represents features common to all DERs of the same type. In contrast, the stochastic parameter characterizes individual variations of DERs' degradation processes, e.g., degradation rates. The error term is included to capture the inherent degradation uncertainties due to signal manifestations and measurement errors. While degradation in DERs may be subjected to time-varying loading as a function of environmental conditions, in this paper we assume that the loading levels remain constant.

It is assumed that the deterministic parameter $\kappa^j$ is known and constant while the stochastic parameter $\phi_{m,i}^j$ follows some distributional form across the population of DERs. An initial estimation of the stochastic degradation parameter distribution, denoted by $\pi(\phi_{m,i}^j)$, can be obtained through the engineering knowledge and historical data related to DERs. The real-time sensor information collected from the DER enables us to update the initial distribution of the stochastic parameter to its posterior distribution counterpart $\tilde{\pi}(\phi_{m,i}^j)$ using the Bayesian updating procedure. 
}We define the remaining life of DER $i$ of type $j$ in $m^{th}$ microgrid at observation time $t_{\mathrm{o}}$, ${R}^{t_{\mathrm{o}},j}_{m,i}$, 
as the first time that the future trajectory of its degradation signal crosses the failure threshold $\theta^{j}_{m,i}$. {\ff Given the updates on the DER's degradation parameter, the remaining life of the DER at the observation time $t_\mathrm{o}$ can be evaluated as: 
\begin{equation}
P\big(R^{t_\mathrm{o},j}_{m,i}=t\big) =P\big(t= \min\big[ s \geq 0 | D^{j}_{m,i}\big(s|\tilde{\pi}(\phi_{m,i}^j)\big) \geq \theta^{j}_{m,i}\big]\big)  
\label{eq:R_update}
\end{equation}
For more details on this class of degradation models see \cite{gebraeel2005residual}.

Given the updates on the RLD of the DERs,} we calculate the expected cost of conducting maintenance $t$ time units after the observation time $t_o$ as follows\cite{yildirim2016sensor,Yildirim2016SensorDrivenCG,yildirim2017integrated}: 
\begin{equation}
  C^{t_o,j}_{m,i,t} = \frac{ C^{p,j}_{m,i} P(R^{t_o,j}_{m,i} > t) + C^{f,j}_{m,i} P(R^{t_o,j}_{m,i} \leq t) }{\int_0^{t} P(R^{t_o,j}_{m,i} > z) dz + t_o} 
\end{equation}
$C^{p,j}_{m,i}$ and $C^{f,j}_{m,i}$ are preventive maintenance (PM) and corrective maintenance (CM) costs, respectively. The function models the trade-off between the risk of unexpected failures and the cost of PM actions by incorporating the corresponding probabilities. We note that dynamic maintenance costs adapt to DERs' health condition, since $R^{t_o,j}_{m,i}$ is updated through the sensor information.
\vspace{-4mm}
\subsection{Adaptive Predictive Operations and Maintenance of MMG}
In this section we formulate a predictive operation and maintenance optimization model that fully adapts to sensor-driven predictions on RLD and dynamic maintenance costs. The proposed  Sensor-Driven Integrated Operations and Maintenance Scheduling Model is a stochastic mixed-integer program that incorporates uncertainties from renewable generation, loads, and market prices.

We study a general setting for an MMG system of multiple interconnected microgrids that can exchange energy with each other, and with the grid. We assume 
each microgrid has several WTs, PVs, CGs, batteries, as well as critical and non-critical local loads.  
The goal is to maintain and operate microgrids in a collaborative manner in order to minimize the total operational and maintenance costs.
We model a complex operational problem that explicitly models the tradeoffs between satisfying flexible local loads, charging battery systems, and trading electricity with other microgrids and the grid. {For trade decisions, we also consider power loss during transfer of electricity.} {\ff We assume that power loss is manifested as a fixed percentage of the electricity flow.} 
Under emergencies, the MMG may also operate in the locally-connected mode, in which only MGs are connected to each other, or islanded mode that isolates each microgrid. {\ff For the $m^{th}$ microgrid, $N(m)$ represents the set of neighbouring microgrids with active connection.}

 {\ff Each MG has two types of DERs: renewable and non-renewable DERs.} Set of renewable  and conventional DERs within the $m^{th}$ microgrid are represented by $\mathcal{G}_{m}^{\mathrm{r}}$, and ${\ff\mathcal{G}_{m}^{\mathrm{nr}}}$, respectively. DERs are further partitioned into two {\ff subsets}: operational and failed. Operational and failed renewable DERs are denoted as $\mathcal{G}_{m}^{\mathrm{r},\mathrm{o}}$, and $\mathcal{G}_{m}^{\mathrm{r},\mathrm{f}}$. {\ff Conventional} DERs follow the same notation. Operational DERs can be scheduled for PM while failed ones can only go under CM actions. For this purpose, we introduce binary variable $\nu_{m,i,t}^{j}$, which determines the start time of PM for the operational DER $i$ of type $j$ within $m^{th}$ microgrid. The dynamic maintenance cost, $C^{t_\mathrm{o},j}_{m,i,t}$ that was introduced in section \ref{section:DegradationAnalytics} corresponds to the cost associated with this decision. We note that the dynamic maintenance cost is computed from the remaining life prediction of the operating DER and is updated based on the most recent sensor observations. 
For failed DERs, binary variable $z^j_{m,i,t}$ represents the start time of CM action. A failed DER cannot dispatch until fixed. Further, the maintenance crew visits are modeled as binary decision variables ${\ff x^\mathrm{crew}_{m,t}}$, which equals to $1$ if maintenance crew visits the microgrid $m$ at time $t$.

The objective is to leverage on the sensor observations to minimize the operations and maintenance costs over MMG:
\vspace{-2mm}

{\ff \begin{equation}\label{eq:OBJ} 
\begin{aligned} \min&\sum_{m=1}^M\Bigg(\underbrace{\sum_{j=1}^J\sum_{i=1}^ {G_{m}^{j,o}}\sum_{t =1}^TC^{t_\mathrm{o},j}_{m,i,t}\cdot \nu^j_{m,i,t}+\sum_{t =1}^TC^{\mathrm{crew}}_{m}\cdot x^\mathrm{crew}_{m,t}}_{\text{Maintenance cost of operational DERs}}\\ &+\sum_{i=1}^ {G_{m}^{nr,o}}\sum _{\omega=1}^{|\Omega|}\sum_{t=1}^T\sum_{h=1}^Hp_{\omega}\big(C^{\mathrm{on}}_{m,i}\cdot \beta^{\mathrm{on},\omega}_{m,i,t,h}+C^{\mathrm{off}}_{m,i}\cdot\beta^{\mathrm{off},\omega}_{m,i,t,h}\\
&\underbrace{+V_{m,i}\cdot x_{m,i,t,h}^{\omega}+B_{m,i}\cdot y_{m,i,t,h}^{\omega}\big)}_{\text{Expected production cost of non-renewable DERs}}\\ 
& \underbrace{+\sum _{\omega=1}^{|\Omega|}\sum_{t=1}^T\sum_{h=1}^Hp_{\omega}\big(\Gamma_{t,h}^{\mathrm{gb},\omega} \cdot y_{m,t,h}^{gb,\omega}-\Gamma_{t,h}^{\mathrm{gs},\omega}\cdot y_{m,t,h}^{\mathrm{gs},\omega}\big)}_{\text{Expected cost of power transaction with the grid}}  \\
& \underbrace{+\sum _{\omega=1}^{|\Omega|}\sum_{t=1}^T\sum_{h=1}^Hp_{\omega}\big(\lambda_m^\mathrm{{c}}\cdot\psi^{\mathrm{c},\omega}_{m,t,h}+\lambda_m^\mathrm{{n}}\cdot\psi^{\mathrm{n},\omega}_{m,t,h}\big)}_{\text{Expected load curtailment cost}}\Bigg) 
\end{aligned}
\end{equation}}
{The objective function evaluates the expected maintenance and operational costs of DERs}. The first term corresponds to the maintenance cost of DERs, including the dynamic maintenance cost of DERs and the crew deployment costs. {\ff The second, third and fourth terms represent the expected operational cost of MMG, which consist of: i) production costs of non-renewable DERs, ii) power transactions cost of microgrids with the grid, and iii) load curtailment costs. The expected hourly production costs of non-renewable DERs include start-up, shut down, and generation costs. Microgrids can sell their excess power to the grid or purchase power from the grid. The associated cost of power transaction with the grid is included in the third term based on the corresponding market price in scenario $\omega$. The MMG loads are prioritized as critical and non-critical. Curtailing critical loads would be penalized harsher than non-critical ones. Finally, the expected hourly load curtailment costs are included in the last part.}
We next introduce the model constraints.
\subsubsection{\textbf{Maintenance Coordination}} 
To guarantee a certain level of generation reliability, constraint (\ref{eq:mc1}) ensures that each DER $i$ of type $j$ is maintained before a dynamic time limit $\varphi_{m,i}^j$. This time limit is defined as the first time that the DER's sensor-updated reliability falls below a predefined control threshold $\Lambda_{m,i}^j$., i.e. $\varphi_{m,i}^j:=min\{t \in \mathcal{T} : P(R_{m,i}^{\mathrm{t_o},j} > t) <\Lambda_{m,i}^j\}$.
\vspace{-3mm}
\begin{equation}\label{eq:mc1}
\sum_{t=1}^{\varphi_{m,i}^j}\nu^j_{m,i,t}=1,\quad \forall m \in \mathcal{M},\forall j \in \mathcal{J},\forall i \in \mathcal{G}^{\mathrm{o},j}_m
\vspace{-2mm}
\end{equation} 
{\ff Constraints \eqref{eq:mcc1} and \eqref{eq:mcc2} ensure the coupling of DERs maintenance decisions and maintenance crew visits.  We assume that PM and CM of DER $i$ take $Y^{\mathrm{o}}_i$ and $Y^{\mathrm{f}}_i$ weeks, respectively. For DER $i$ which goes under PM at time $t$; $\nu^j_{m,i,t}=1$, constraint \eqref{eq:mcc1} guarantees the maintenance crew presence during the maintenance time $k \in \{t,..,t+Y^{\mathrm{o}}_i\}.$ The same logic applies for constraint \eqref{eq:mcc2}.} In addition, constraint \eqref{eq:mcc3} enforces that a maintenance crew cannot visit multiple microgrid locations simultaneously. 
\vspace{-2.5mm}
\begin{align}
&\sum_{k=0}^{Y^{\mathrm{o}}_i-1}\nu^j_{m,i,t-k}\leq {\ff x^\mathrm{crew}_{m,t}},\mkern3mu  \forall m \in \mathcal{M},\forall j \in \mathcal{J}, \forall i \in \mathcal{G}^{j,\mathrm{o}}_m, \forall t \in \mathcal{T}\label{eq:mcc1}\raisetag{11pt}
\\
&\sum_{k=0}^{Y^{\mathrm{f}}_i-1}z^j_{m,i,t-k}\leq {\ff x^\mathrm{crew}_{m,t}},\mkern3mu  \forall m \in \mathcal{M}, \forall j \in \mathcal{J}, \forall i \in \mathcal{G}^{j,\mathrm{f}}_m, \forall t \in \mathcal{T}\raisetag{11pt}\label{eq:mcc2}\\
&\sum_{m \in \mathcal{M}}{\ff x^\mathrm{crew}_{m,t}}\leq 1,\quad\forall t \in \mathcal{T}\label{eq:mcc3}
\vspace{-2.3mm}
\end{align}
\subsubsection{\textbf{Maintenance \& Operations Coupling}} 
The maintenance decision variables $z^j_{m,i,t}$ {\ff and} $\nu^j_{m,i,t}$ are coupled with the dispatch decisions $y_{m,i,t,h}^{j,\omega}$. Non-renewable DERs typically have operational limitations such as maximum and minimum generation levels, while the renewable DERs mainly work at their maximum power point. This small operational difference causes slight variation in modeling.

\emph{Renewable DERs:} Constraint \eqref{eq:mc2} ensures that i) operational renewable DERs produce electricity within their available capacity, namely $\Phi_{m,i,t,h}^{\omega}$, which depends on the solar or wind power availability at scenario $\omega$, week $t${\ff, and} hour $h$; and ii) units under maintenance can not produce electricity {\ff in any of the hours within the maintenance periods.}
\vspace{-2mm}
\begin{equation}
\begin{aligned}\label{eq:mc2}
&\qquad \qquad y_{m,i,t,h}^{\omega}\leq \Phi_{m,i,t,h}^{\omega}(1-\sum_{k=0}^{Y^{o}_i-1}\nu^{\mathrm{r}}_{m,i,t-k})\qquad\\&\forall m \in \mathcal{M},\forall i \in \mathcal{G}^{r,\mathrm{o}}_m,\forall t \in \mathcal{T},\forall h \in \mathcal{H},\forall \omega \in \Omega
\end{aligned}
\end{equation}
{\ff Constraint \eqref{eq:CM} guarantees two factors. Firstly, a renewable DER that started at a failed state cannot produce electricity until it undergoes CM for $Y_i^{\mathrm{f}}$ weeks, i.e. $z^{\mathrm{r}}_{m,i,k}=1$ for any week $k \in \{1,..,t-Y_i^{\mathrm{f}}\}$. Secondly, a correctively maintained DER can produce electricity up to its available capacity.}
\vspace{-3mm}
\begin{equation}
\begin{aligned}\label{eq:CM}
&\qquad \qquad y_{m,i,t,h}^{\omega}\leq \Phi_{m,i,t,h}^{\omega}\sum_{k=1}^{t-Y^{\mathrm{f}}_i}z^{\mathrm{r}}_{m,i,k}\qquad\\&\forall m \in \mathcal{M},\forall i \in   \mathcal{G}^{\mathrm{r},\mathrm{f}}_m,\forall t \in \mathcal{T},\forall h \in \mathcal{H},\forall \omega \in \Omega
\end{aligned}
\end{equation}

\emph{Non-renewable DERs:} Constraint (\ref{eq:mc4}) couples the PM decision variables $\nu^j_{m,i,t}$ with the corresponding commitment variable $x_{m,i,t,h}^\omega$ for non-renewable DERs. This constraint ensures that if a unit $i$ is under maintenance during week $t$, it cannot be committed within that period.
\vspace{-2mm}
\begin{equation}
\begin{aligned}\label{eq:mc4}
&\qquad \qquad \quad x_{m,i,t,h}^\omega\leq 1-\sum_{k=0}^{Y^{o}_i-1}\nu^{\mathrm{nr}}_{m,i,t-k}\quad \\ &\forall m \in \mathcal{M}, \forall i \in \mathcal{G}^{\mathrm{nr},\mathrm{o}}_m,\forall t \in \mathcal{T},\forall h \in H,\forall \omega \in \Omega\\ 
\end{aligned}
\end{equation}
Constraint \eqref{eq:mc3} enforces that a failed unit should be scheduled for a CM before it can generate electricity.
\vspace{-3mm}
\begin{equation}
\begin{aligned}\label{eq:mc3}
&\qquad \qquad x_{m,i,t,h}^\omega\leq \sum_{k=1}^{t-Y^{f}_i}z^{\mathrm{nr}}_{m,i,k}\\&\forall m \in \mathcal{M},\forall i \in  \mathcal{G}^{\mathrm{nr},\mathrm{f}}_m,\forall t \in \mathcal{T},\forall h \in \mathcal{H},\forall \omega \in \Omega
\end{aligned}
\end{equation}
\subsubsection{\textbf{Load Management}}
The critical and non-critical loads of $m^{th}$ microgrid are represented by $D^{\mathrm{c},\omega}_{m,t,h}$ and $D^{\mathrm{n},\omega}_{m,t,h}$, respectively. 
Constraints \eqref{eq:LoadC} {\ff and }\eqref{eq:LoadNC} ensure that for each type of load, curtailed load does not not exceed the total load.
\begin{align}
\hspace{-3mm} \psi^{\mathrm{c},\omega}_{m,t,h}\leq D^{\mathrm{c},\omega}_{m,t,h},\quad  \forall m \in \mathcal{M}, \forall t \in \mathcal{T},\forall h \in \mathcal{H}, \forall \omega \in \Omega\label{eq:LoadC}\\
\hspace{-3mm}
\psi^{\mathrm{n},\omega}_{m,t,h} \leq D^{\mathrm{n},\omega}_{m,t,h} ,\quad \forall m \in \mathcal{M},\forall t \in \mathcal{T}, \forall h \in \mathcal{H}, \forall \omega \in \Omega\label{eq:LoadNC}
\end{align}
\subsubsection{\textbf{Power Transactions}} 
Each microgrid has a bidirectional power flow with the grid and {\ff neighboring} microgrids.
Binary variables $g_{m,t,h}^{\mathrm{p},\omega}$ and $g_{m,t,h}^{\mathrm{s},\omega}$ describe the power exchange status of $m^{th}$ microgrid with the grid: i.e.{\ff ,} buying and selling.  Constraint \eqref{eq:bsGrid} determines the direction of flow between {\ff the} $m^{th}$ microgrid and the grid. We represent the microgrid purchased power from the main grid and sold power to the main grid with $y_{m,t,h}^{\mathrm{gp},\omega}$, and $y_{m,t,h}^{\mathrm{gs},\omega}$, respectively. {\ff Purchasing and selling limits, $\overline{F_{m}^{\mathrm{gp}}}$ and $\overline{F_{m}^{\mathrm{gs}}}$, are enforced through constraint \eqref{eq:GridL}.} 
MMG in locally connected mode can be modeled by setting $g_{m,t,h}^{\mathrm{p},\omega}$ and $g_{m,t,h}^{\mathrm{s},\omega}$ to zero. 
\begin{align}
&g_{m,t,h}^{\mathrm{p},\omega}+g_{m,t,h}^{\mathrm{s},\omega}\leq 1\label{eq:bsGrid}\\
&y_{m,t,h}^{\mathrm{gp},\omega}\leq g_{m,t,h}^{\mathrm{p},\omega} \cdot \overline{F_{m}^{\mathrm{gp}}}, \quad y_{m,t,h}^{\mathrm{gs},\omega} \leq  g_{m,t,h}^{\mathrm{s},\omega} \cdot \overline{F_{m}^{\mathrm{gs}}}\label{eq:GridL}\\ \nonumber
&\qquad \forall m \in \mathcal{M},\forall t \in \mathcal{T},\forall h \in \mathcal{H},\forall \omega \in \Omega
\end{align}
Likewise, $u_{m,l,t,h}^{\mathrm{p},\omega}$ and $u_{m,l,t,h}^{\mathrm{s},\omega}$ indicate the power transaction status of $m^{th}$ microgrid with the neighbouring microgrids $l \in N(m)$. Purchased power from the neighbouring microgrid $l$, and the power sold to microgrid $l$ are represented by $y_{m,l,t,h}^{\mathrm{p},\omega}$ and $y_{m,l,t,h}^{\mathrm{s},\omega}$, respectively. Constraint \eqref{eq:MGsEx} guarantees that $m^{th}$ microgrid cannot simultaneously buy and sell power from its neighbouring microgrid $l$. Constraints \eqref{eq:MGsExUS} enforce flow limits {\ff on power transactions between microgrids. These constraints guarantee that the purchased and sold power between two neighboring microgrids do not exceed their corresponding limitations $\overline{F_{m,l}^{\mathrm{p}}}$ and $\overline{F_{m,l}^{\mathrm{s}}}$}.
\begin{align}
&u_{m,l,t,h}^{\mathrm{p},\omega}+u_{m,l,t,h}^{\mathrm{s},\omega} \leq 1 \label{eq:MGsEx},\\
&y_{m,l,t,h}^{\mathrm{p},\omega} \leq u_{m,l,t,h}^{\mathrm{p},\omega} \cdot \overline{F_{m,l}^{\mathrm{p}}}, \quad y_{m,l,t,h}^{\mathrm{s},\omega} \leq u_{m,l,t,h}^{\mathrm{s},\omega} \cdot \overline{F_{m,l}^{\mathrm{s}}}\label{eq:MGsExUS},\\\nonumber
&\forall m \in \mathcal{M},\forall l \in N(m),\forall  t \in \mathcal{T},\forall h \in \mathcal{H},\forall \omega \in \Omega
\end{align}
 We can model the islanded microgrids mode by limiting electricity flow across microgrids, $g_{m,t,h}^{\mathrm{p},\omega}=g_{m,t,h}^{\mathrm{s},\omega}=u_{m,l,t,h}^{\mathrm{p},\omega}=u_{m,l,t,h}^{\mathrm{s},\omega} =0, \forall l \in N(m)$. We also enforce constraints ensuring flow of electricity across microgrids and with the grid in presence of power dissipation.

\subsubsection{\textbf{Storage Operation}} 
In constraint \eqref{eq:SOC}, the state of charge (SOC) of $b^{th}$ battery within $m^{th}$ microgrid, at week $t$ and hour $h$, is coupled with its previous SOC, the charged and discharged power during that time, and the battery charging/discharging efficiency, $\eta_{m,b}$. The battery's SOC cannot exceed the maximum capacity and cannot reduce below the manufacturer recommended value. The maximum and minimum SOC is limited by (\ref{eq:ESDstatus}). We denote the charging and discharging status of battery by $e_{m,b,t,h}^{\textbf{+},\omega}$ and $e_{m,b,t,h}^{\textbf{-},\omega}$, respectively. Simultaneous charging and discharging is not possible and is guaranteed through constraint (\ref{eq:ESDstatus}). 
In addition, batteries charging-discharging amount is limited to their rated power capacity and is modeled through constraint \eqref{eq:bdch}.
\vspace{-1mm}
\begin{align}
&soc^{\omega}_{m,b,t,h}=soc^{\omega}_{m,b,t,h-1}+\eta_{m,b}\cdot \pi^{\textbf{+},\omega}_{m,b,t,h}-\frac{\pi^{\textbf{--},\omega}_{m,b,t,h}}{\eta_{m,b}}\label{eq:SOC}\\
&\underline{P^{soc}_{m,b}}\leq soc^{\omega}_{m,b,t,h}\leq \overline{P^{soc}_{m,b}}, \quad e_{m,b,t,h}^{\textbf{+},\omega}+e_{m,b,t,h}^{\textbf{-},\omega}\leq 1 \label{eq:ESDstatus}\\
&\pi^{\textbf{+},\omega}_{m,b,t,h}\leq e_{m,b,t,h}^{\textbf{+},\omega}\cdot \overline{P_{m,b}^{ch}}, \quad  \pi^{\textbf{--},\omega}_{m,b,t,h}\leq e_{m,b,t,h}^{\textbf{-},\omega} \cdot \overline{P_{m,b}^{dch}}\label{eq:bdch}\\& \qquad\quad\forall m \in \mathcal{M},\forall b \in \mathcal{B}_m,\forall t \in \mathcal{T},\forall h \in \mathcal{H},\forall \omega \in \Omega\nonumber
\end{align}

We also enforce that the battery SOC remains the same at the first and last hour of each week $t$.

\subsubsection{\textbf{Power Balance}} 
Constraint (\ref{eq:powerbalance}) models the power balance of $m^{th}$ microgrid. The left-hand side denotes the microgrid's purchased power from the grid, purchased power from neighboring microgrids, DERs generation, and batteries discharged power, respectively. The right-hand side includes sold power to the grid and other microgrids, the net non-critical and critical loads, and charged power to batteries, respectively.

\begin{align}\label{eq:powerbalance}
&y_{m,t,h}^{\mathrm{gp},\omega}+\sum_{\substack{l\in N(m)}}y_{m,l,t,h}^{\mathrm{p},\omega}+\sum_{j\in \mathcal{J}}\sum_{i\in \mathcal{G}_{m}^{j}}y_{m,i,t,h}^{\omega}+\sum_{b\in \mathcal{B}_m}\pi_{m,b,t,h}^{\textbf{--},\omega}=\nonumber\\&y_{m,t,h}^{\mathrm{gs},\omega}+\sum_{\substack{l\in N(m)}}y_{m,l,t,h}^{\mathrm{s},\omega}+D_{m,t,h}^{\mathrm{n},\omega}-\psi^{\mathrm{n},\omega}_{m,t,h}+D_{m,t,h}^{\mathrm{c},\omega}-\psi^{\mathrm{c},\omega}_{m,t,h}\nonumber\\ & +\sum_{b\in \mathcal{B}_m}\pi_{m,b,t,h}^{\textbf{+},\omega}, \; \forall m \in \mathcal{M},   \forall t \in \mathcal{T},  \forall h \in \mathcal{H},  \forall \omega \in \Omega
\end{align}
\subsubsection{\textbf{Non-Renewable DER Operation}} 
{\ff For conventional generators, binary variable $x_{m,i,t,h}^\omega$ represents the hourly commitment decisions while $\beta^{\mathrm{on},\omega}_{m,i,t,h}$ and $\beta^{\mathrm{off},\omega}_{m,i,t,h}$ indicate start-up and shut-down decisions, respectively. Constraints (\ref{eq:Genon}) and (\ref{eq:Genoff}) represent the logic relations between on and off status of generators and turn on and turn off actions.
\begin{align}
&x^{\omega}_{m,i,t,h-1}-x^{\omega}_{m,i,t,h}+\beta^{\mathrm{on},\omega}_{m,i,t,h}\geq 0\label{eq:Genon}\\
&x^{\omega}_{m,i,t,h}-x^{\omega}_{m,i,t,h-1}+\beta^{\mathrm{off},\omega}_{m,i,t,h}\geq 0\label{eq:Genoff}
\\&\nonumber\qquad\qquad\qquad\quad \forall m \in \mathcal{M},\forall i \in \mathcal{G}^{\mathrm{nr}}_m,\forall t \in \mathcal{T},  \forall h \in \mathcal{H},  \forall \omega \in \Omega
\end{align}
Constraints (\ref{eq:GenMinU}) and (\ref{eq:GenMinD}) represent the minimum up and minimum downtime of conventional DERs. If conventional DER $i$ is turned on at hour $h$, it must remain on at least for the next $MU_{m,i}$ hours. The same logic is valid for the minimum downtime constraint.
\begin{align}
&x^{\omega}_{m,i,t,h-1}-x^{\omega}_{m,i,t,h}\leq x^{\omega}_{m,i,t,h^{'}},\nonumber\\& \forall h^{'}\in[h+1,\min\{h+MU_{m,i}-1,H\}],h\in [2,H],\label{eq:GenMinU}\\
&x^{\omega}_{m,i,t,h}-x^{\omega}_{m,i,t,h-1}\leq 1-x^{\omega}_{m,i,t,h^{'}},\nonumber\\& \forall h^{'}\in[h+1,\min\{h+MD_{m,i}-1,H\}],h\in [2,H],\label{eq:GenMinD}\\&\nonumber\qquad \forall m \in \mathcal{M},\forall i \in \mathcal{G}^{\mathrm{nr}}_m,\forall t \in \mathcal{T},  \forall \omega \in \Omega
\end{align}
The continuous variable  $y^{\omega}_{m,i,t,h}$ represents the generation output of non-renewable DERs at each hour $h$ within week $t$ under scenario $\omega$.  Generation output of committed non-renewable DERs is limited to their maximum and minimum production capacity through constraint (\ref{eq:GenCap}). Constraint (\ref{eq:GenRamp}) is the ramping constraint.
\begin{align}
&\underline{P^{\mathrm{nr}}_{m,i}}\cdot x^{\omega}_{m,i,t,h}\leq y^{\omega}_{m,i,t,h}\leq \overline{P^{\mathrm{nr}}_{m,i}}\cdot x^{\omega}_{m,i,t,h},\label{eq:GenCap}\\
&-RD_{m,i}\leq y^{\omega}_{m,i,t,h}-y^{\omega}_{m,i,t,h-1}\leq RU_{m,i},\label{eq:GenRamp}
\\&\nonumber\qquad\qquad\qquad \forall m \in \mathcal{M},\forall i \in \mathcal{G}^{\mathrm{nr}}_m,\forall t \in \mathcal{T},  \forall h \in \mathcal{H},  \forall \omega \in \Omega
\end{align}}
\vspace{-3mm}
\section{Two-Stage Stochastic Reformulation}\label{section:reformulation}
Joint operations and maintenance problems belong to a class of computationally demanding problems due to their size and model complexity. Integration of sensor information adds an additional layer of difficulty to this challenge. This section introduces a two-stage reformulation of SD-IOM to motivate an iterative solution algorithm. To introduce our algorithm, we first define the compact matrix formulation of the SD-IOM model - deterministic equivalent model - as follows:
\begin{subequations}  \label{eq:(SD-IOM)}
	\begin{align}
	\min \quad\boldsymbol{a}^\top \boldsymbol{ z } +\sum_{\omega \in \Omega}p_{\omega}\big(\boldsymbol{q}^\top \boldsymbol{x}_{\omega}+\boldsymbol{b}_\omega^\top \boldsymbol{y}_{\omega}\big) \\ \label{eq:(SD-IOM)1.2}
	\text{s.t.}  \quad\boldsymbol{A}\boldsymbol{ z }\textcolor{white}{p\sum p_{\omega}\qquad\big(L\boldsymbol{x}_{\omega}+G\boldsymbol{y}_{\omega}\big)} &\leq \boldsymbol{g}\\ \label{eq:(SD-IOM)1.3}
	\boldsymbol{H}_{\omega}\boldsymbol{z} +\textcolor{white}{p p_{\omega}\big(p_{\omega}\big(p}\boldsymbol{E}\boldsymbol{ x }_{\omega}+\boldsymbol{D}\boldsymbol{y}_{\omega}\textcolor{white}{p }&\leq \boldsymbol{e}\\ \label{eq:(SD-IOM)1.5}
    \boldsymbol{L}\boldsymbol{ x }_{\omega}   +  \boldsymbol{G y }_{\omega}\textcolor{white}{p} &\leq \boldsymbol{\ell }_\omega
	\end{align}
	\vspace{-0.7cm}
    \begin{align}
    	{\ff\qquad\boldsymbol{x}_{\omega} \in \{ 0,1 \}^{\big(3G^{\mathrm{nr}}+2B+2 M+\sum\limits_{m=1}^{M}2N(m)\big)\cdot T \cdot H\cdot |\Omega|}}\nonumber\\
        	 {\ff\boldsymbol{z}\in \{ 0,1 \}^{M\cdot T+G\cdot T},}\boldsymbol{y}_{\omega} \geq 0, \forall \omega \in \Omega\hspace{1.2cm}\nonumber
\end{align}
\end{subequations}

{\ff where $\boldsymbol{z}$ represents the maintenance-related decision variables including the PM/CM maintenance and crew visit decisions.} $\boldsymbol{x}_{\omega}$ and  $\boldsymbol{y}_{\omega}$ are binary and continuous operational decision variables associated with the optimal power dispatch of the MMG. In this form, constraint  \eqref{eq:(SD-IOM)1.2} corresponds to maintenance actions, such as maintenance crew capacity constraints and PM/CM coordination. Constraint \eqref{eq:(SD-IOM)1.3} couples maintenance decisions with DERs operational decisions so that under maintenance or failed DERs cannot generate power. Other operational constraints such as DERs production specifications, power transaction of microgrids, etc are represented by equation \eqref{eq:(SD-IOM)1.5}.

The SD-IOM model \eqref{eq:(SD-IOM)} is naturally formulated as a two-stage stochastic program with complete recourse, in which maintenance decisions are determined in the first stage, and operational decisions given the maintenance schedules reside in the second-stage, as follows: 
\vspace{-1mm}
\begin{subequations} 
	\begin{align}
	\underset{\boldsymbol{z}}{\text{min}} \hspace{2mm}\boldsymbol{a}^\top \boldsymbol{ z } +\sum_{\omega \in \Omega} p_{\omega}Q_\omega(\boldsymbol{z})\textcolor{white}{p\sum p_{\omega}\qquad} \\ 
	\text{s.t.}  \quad                  \boldsymbol{A z }       \leq \boldsymbol{ g}\hspace{3cm}\\
	{\ff\boldsymbol{z}\in \{ 0,1 \}^{M\cdot T+G\cdot T}\hspace{1.05cm}\nonumber}
	\end{align} 
	\label{eq:AOMOIIM} 
\end{subequations} 
$Q_\omega(\boldsymbol{z})$ denotes the second-stage operational scheduling problem given maintenance decision $\boldsymbol{z}$ under the realized scenario $\omega$. Without loss of generality, minimum up/down, and ramping constraints of CGs are considered within the hours of the same week. 
Consequently, once the maintenance decisions are determined, the operational decisions for any scenario $\omega$ and week $t$ become independent. So, the second-stage operational problem $Q_\omega(\boldsymbol{z})$ is equal to $\sum_{t}Q_{t,\omega}(\boldsymbol{z}_t)$, where $\forall \omega \in \Omega$ and $\forall t \in \mathcal{T}$, $Q_{t,\omega}(\boldsymbol{z}_t)$ is as follows:
\begin{subequations} 
\begin{align}
\text{min}\hspace{2mm}  \boldsymbol{q}_t^\top \boldsymbol{x}_{t,\omega}+\boldsymbol{b}_{t,\omega}^\top \boldsymbol{y}_{t,\omega}\hspace{2.2cm}\\ 
\hspace{1.7cm}\text{s.t.}\hspace{1mm} \boldsymbol{E}_t\boldsymbol{x}_{t,\omega} +\boldsymbol{D}_t\boldsymbol{y}_{t,\omega}\leq \boldsymbol{e }_t-\boldsymbol{H}_{t,\omega}\boldsymbol{z}_t\\ 
\boldsymbol{L}_t\boldsymbol{x}_{t,\omega}+\boldsymbol{G}_t\boldsymbol{y}_{t,\omega}\leq \boldsymbol{\ell}_{t,\omega}\hspace{1.2cm}
\end{align}
\vspace{-0.7cm}
\begin{align}
{\ff\quad\boldsymbol{x}_{t,\omega} \in \{ 0,1 \}^{\big(3 G^{\mathrm{nr}}+2 B+2 M+\sum\limits_{m=1}^{M}2N(m)\big) \cdot H}}\nonumber\nonumber,\boldsymbol{y}_{t,\omega} \geq 0\nonumber
\end{align}
\end{subequations}
 
{\ff This reformulation approach enables us to accelerate the computational performance in solving  large  scale  problems. Exploiting  the  sub-problems’  operational  independency  per-week  per-scenario  allows  us  to  solve  smaller  sub-problems  in  comparison  with the  per-week  decomposition. In general, adding disaggregate cuts through multi-cut method can decrease the number of major iterations\cite{fazeli2020two}.}

To leverage on these reformulations, we iteratively add \emph{L-Shaped} \cite{van1969shaped} and integer cuts. We add L-shaped optimality cuts either: i) per week, or ii) per scenario and week. 
To apply the method, we first need to relax the operational sub-problems by relaxing the binary operational variables, i.e. $\boldsymbol{x}_{t,\omega} \in [0,1], \forall t \in \mathcal{T}, \forall \omega \in \Omega$. {\ff We represent the objective of the relaxed operational sub-problems by $\mathcal{R}(.)$. } The solution of the relaxed SD-IOM provides a lower bound for the original model \eqref{eq:(SD-IOM)}. Hence, after the L-Shaped method convergence, we add cost recovery cuts (integer cuts) to retrieve the exact operational costs. Algorithms based on a combination of Benders' and integer cuts have been proposed for various problem settings \cite{laporte1993integer, Yildirim2016SensorDrivenCG,rahmaniani2017benders}. In the followings, we discuss the algorithms procedure.

\begin{figure}
  \centering 
    \includegraphics[width=0.44\textwidth]{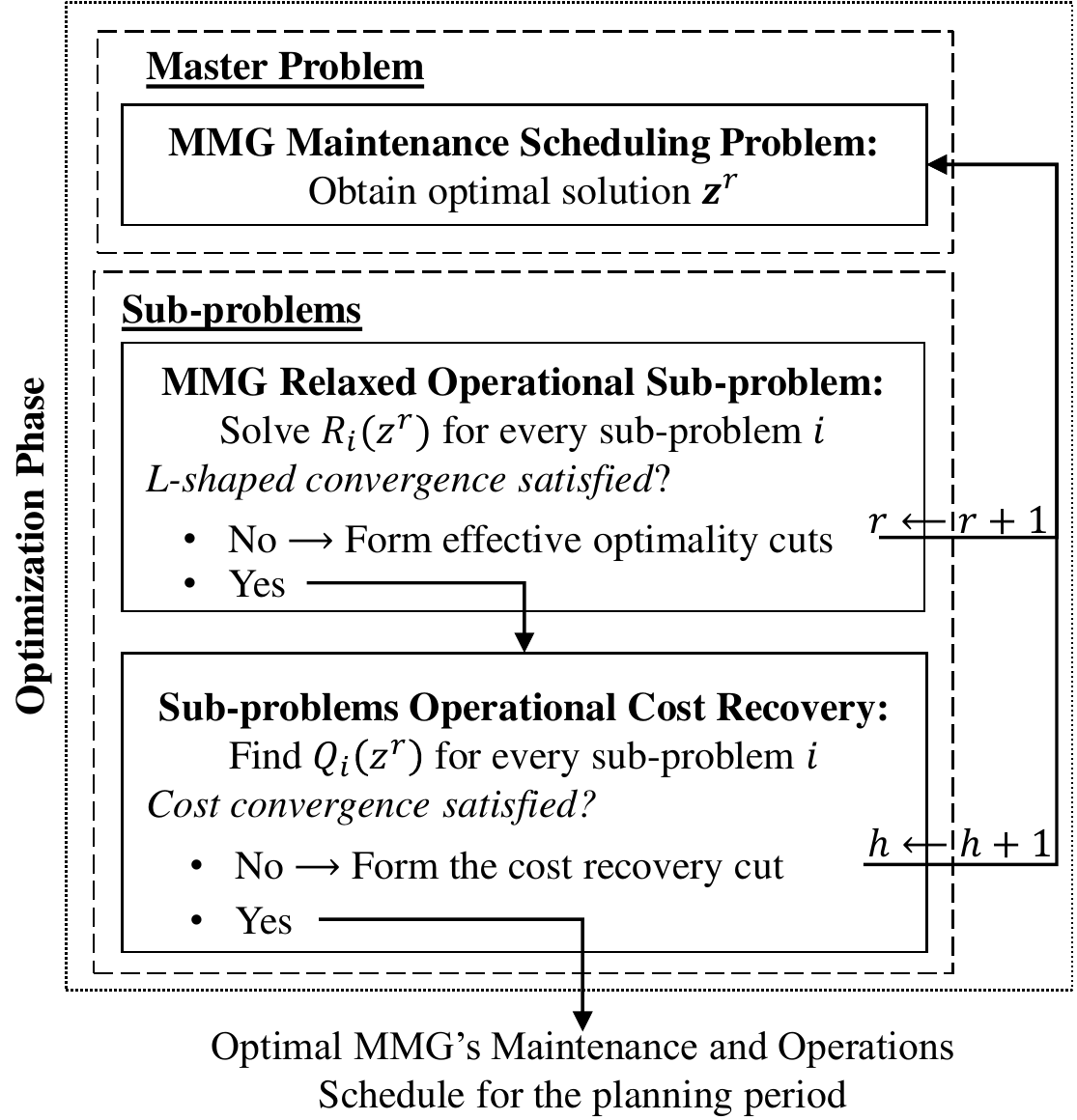} 
    \vspace{1mm}
    \caption{{\ff Flowchart of the proposed algorithms for SD-IOM.} }
    \label{fig:flowchart} 
\end{figure}

{\ff We provide a nested algorithm to solve the SD-IOM problem (outlined in Algorithms 1 and 2). At the outer level, we solve the relaxed SD-IOM problem leveraging the per-week or per-week per-scenario decomposition methods. In the inner level, we solve the cost recovery algorithm. We define $\text{Conv}^\ell$ and $\text{Conv}^i$ as the outer and inner level convergence flags and set them to false. The $\text{UB}^\ell$ and $\text{LB}^\ell$ represent the obtained upper and lower bounds.  We set the first-stage objective function to $\boldsymbol{a}^\top \boldsymbol{ z }+\boldsymbol{\eta}+\theta$.}
The free variable $\boldsymbol{\eta}$ approximates the relaxed second-stage sub-problems {\ff $\mathcal{R}_t(\boldsymbol{z}_t),\forall t \in \mathcal{T}$} through optimality cut \eqref{eq:cutW} or {\ff  sub-problems $\mathcal{R}_{t,\omega}(\boldsymbol{z}_t), \forall t \in \mathcal{T},\forall \omega\in\Omega$} through \eqref{eq:cutWS} {\ff according to Lemma \ref{lemm1}. The free variable $\theta$ recovers the true cost of operations. The per-week algorithm sequentially adds optimality cuts until the convergence criterion is satisfied for the relaxed version of the subproblem (i.e. lower bound). Upon convergence, the cost recovery algorithm (CRA) is executed to incorporate the true cost of operational problems. For each week $t$, we specify DERs' availability based on the obtained optimal maintenance schedule, i.e. $\boldsymbol{ z }^r$. This schedule enabes the evaluation of the exact operational cost $Q_t(\boldsymbol{z}^r_t)$ for each week $t$ in the per week decomposition. 
We then calculate the difference between the operational costs of the exact, $Q^h$, and relaxed, $\mathcal{R}^h$, sub-problems.  The exact and relaxed values are checked to determine if they are close enough and meet the convergence criterion. If the convergence criterion is violated, cost recovery cut is added to the master problem to recover the violated costs, and the algorithm is rerun until the relaxed operational costs and exact operational costs are within a specified convergence limit.  We discuss the L-Shaped per-week and cost recovery cut algorithms in detail in Algorithms 1 and 2. The flowchart in figure \ref{fig:flowchart} illustrates an overall summary of these steps. The multi-cuts per-week per-scenario algorithm follows a similar procedure with minor modifications.}
\begin{figure}[!t]
{\begin{algorithm}[H] 
{
\SetAlgoLined
\caption{Per-week Algorithm}
\label{alg1}
\begin{algorithmic}[1]
\SetNoFillComment
\STATE \textbf{Initialize:} $\text{Conv}^i=\FALSE$, $h=0$.\\ $UB^\ell \leftarrow \infty$, $LB^\ell\leftarrow-\infty$,$\text{Conv}^\ell=\FALSE$, $r=k=0$.\\ 
Define free variables $\eta_t,\forall t \in\mathcal{T}$ and $\theta$.\\
Set $\eta_t^1\leftarrow-\infty,\alpha_t^1\leftarrow0,\beta_t^1\leftarrow0,\forall t \in\mathcal{T}$.\\
Set $\Delta^1\leftarrow0$
\WHILE{$\text{Conv}^i=\FALSE$}
\STATE $h\leftarrow h+1$, $\mathcal{R}^h\leftarrow0$, and  $S^h\leftarrow\emptyset$\\
\WHILE{$\text{Conv}^\ell=\FALSE$ }
\STATE $r\leftarrow r+1$
\STATE Solve first-stage problem
\begin{subequations} 
	\begin{align}
	\underset{\boldsymbol{z}}{\text{min}} \hspace{2mm}\boldsymbol{a}^\top \boldsymbol{ z } +\sum_{t \in \mathcal{T}}\eta_t+\theta\nonumber\hspace{4.3cm} \\ 
	\text{s.t.}  \hspace{1mm}  \boldsymbol{A z }       \leq \boldsymbol{ g}\hspace{5.6cm}\nonumber\\
	\eta_t\geq\alpha_t^k-\boldsymbol{\beta}_t^k\boldsymbol{z}_t\hspace{2mm},\hspace{3mm} \forall t \in \mathcal{T}\nonumber\hspace{2.6cm}\\
	\theta\geq\Delta^h\Phi(\boldsymbol{z},S^h)-\Delta^h(|S^h|-1)\nonumber\hspace{1.7cm}\\
	\boldsymbol{z}\in \{ 0,1 \}^{M\cdot T+G\cdot T},\boldsymbol{\eta}\in R,\theta\in R \hspace{2cm}\nonumber
	\end{align} 
\end{subequations} 
\STATE $LB^\ell\leftarrow \max\{LB^\ell,\boldsymbol{a}^\top \boldsymbol{ z }^r+\sum_{t \in \mathcal{T}}\eta_t^r+\theta\}$
\FORALL{$t \in \mathcal{T}$}
\STATE Solve the relaxed sub-problem $\mathcal{R}_t(\boldsymbol{z}_t^r)$ to obtain dual multipliers $\bm{\pi}^{i,r}_{t,\omega}\hspace{1mm},\hspace{1mm}i=1,2$ and objective value $\mathcal{R}_t(\boldsymbol{z}_t^r)$.
\STATE $\mathcal{R}^h\leftarrow\mathcal{R}^h+\mathcal{R}_t(\boldsymbol{z}_t^r)$
\ENDFOR
\STATE  $UB^\ell\leftarrow \min\{UB^\ell,\boldsymbol{a}^\top \boldsymbol{ z }^r+\sum_{t\in\mathcal{T}}\mathcal{R}_{t}(\boldsymbol{z}_t^r)\}$\\
\IF{$UB^\ell-LB^\ell>\epsilon^\ell|LB^\ell|$}{
\STATE Generate optimality cuts:
\FORALL{$t \in \mathcal{T}$}
\STATE  Compute $\alpha_t$ and $\boldsymbol{\beta}_t$ based on constraint \eqref{eq:Wcut}\\
\IF{$\eta_t^{r}<\alpha_t-\boldsymbol{\beta}_t\boldsymbol{z}^r_t$}{
\STATE $k\leftarrow k+1$
\STATE Set $\alpha_t^k\leftarrow \alpha_t$ , $\boldsymbol{\beta}_t^k\leftarrow \boldsymbol{\beta}_t$}
\ENDIF
\ENDFOR}
\ELSE
\STATE{$\text{Conv}^\ell=\TRUE$}
\ENDIF
\ENDWHILE\\
\tcc{cost recovery algorithm}
\STATE \textbf{Run} $\text{CRA}(\text{Conv}^i,\text{Conv}^\ell,\mathcal{R}^h,S^h,\boldsymbol{z}^r,k)$
\ENDWHILE
\STATE $\boldsymbol{z}*\leftarrow \boldsymbol{z}^r$\\
\KwOut{Optimal Maintenance Schedule $\{ \boldsymbol{ z}^*\}$.}
\end{algorithmic}}
\end{algorithm}}
\end{figure}
{\ff\begin{figure}[!t]
{\ff\begin{algorithm}[H]
\caption{Cost Recovery Algorithm (CRA)}
\label{alg3}
\begin{algorithmic}[1]
\REQUIRE $(\text{Conv}^i,\text{Conv}^\ell,\mathcal{R}^h,S^h,\boldsymbol{z}^r,k)$\\
\STATE \textbf{Initialize:} $Q^h\leftarrow0$, $\Delta^h\leftarrow0$.\\
Define $S^h=\{i|\boldsymbol{z}_i^r=1\}$
\FOR{$t \in \mathcal{T}$}
\STATE Solve $Q_t(\boldsymbol{z}^r)$
\STATE $Q^h\leftarrow Q^h+Q_t(\boldsymbol{z}^r)$
\STATE $\delta_t^h \leftarrow Q_t(\boldsymbol{z}_t^r) - \mathcal{R}_t(\boldsymbol{z}_t^r)$
\ENDFOR
\IF{ $\mathcal{R}^h-Q^h > |Q^h|\epsilon^c$}
\STATE $\text{Conv}^i=\FALSE$ , $\text{Conv}^\ell=\FALSE$
\STATE $k\leftarrow k+1$
\STATE $S^h\leftarrow S^h\cup\{i|\boldsymbol{z}_i^r=1\}$ ,  $\Delta^h\leftarrow\sum_{t\in\mathcal{T}}\delta^h_t$
\STATE$\Phi(\boldsymbol{z},S^h):=\sum\limits_{i\in S^h}\boldsymbol{z}_i-\sum\limits_{i\notin S^h}\boldsymbol{z}_i$
\ELSE
\STATE {$\text{Conv}^i=\TRUE$}
\ENDIF
\ENSURE  $(\text{Conv}^i,\text{Conv}^\ell,\Phi(\boldsymbol{z},S^h),\Delta^h,k)$
\end{algorithmic}
\end{algorithm}}
\end{figure}}
{\ff\begin{lemma} \label{lemm1}
Let  $\bm{\pi}^{1,r}_{t,\omega}$ and $\bm{\pi}^{2,r}_{t,\omega}$ denote the dual multipliers associated with the optimal solution of the sub-problem for week $t$  and scenario $\omega$ at iteration $r$. Then:

Constraints \eqref{eq:cutW} represents the optimality cuts in the per week decomposition method.

\begin{align}\label{eq:cutW}
&\eta_{t}\geq\alpha_t-\boldsymbol{\beta}_t \boldsymbol{z}_t\hspace{1mm},\hspace{2mm}\forall t \in \mathcal{T}
\end{align}
where $\alpha_t$ and  $\boldsymbol{\beta}_t$ are defined as :

\begin{align}\label{eq:Wcut}
&\alpha_t=\sum\limits_{\omega\in\Omega}\big((\pi_{t,\omega}^{1,r})^\top e_t+(\pi_{t,\omega}^{2,r})^\top \ell_{t,\omega}\big)\hspace{1mm},\hspace{1mm}
\boldsymbol{\beta}_t=\sum\limits_{\omega\in\Omega}(\pi_{t,\omega}^{1,r})^\top H_{t,\omega}
\end{align}

Constraint \eqref{eq:cutWS} represents the optimality cuts in the per week per scenario decomposition method.
\begin{align}
\label{eq:cutWS}
&\eta_{\omega,t}\geq \alpha_{t,\omega}- \boldsymbol{\beta}_{t,\omega}\boldsymbol{z}_t\hspace{1mm},\hspace{2mm}\forall \omega \in \Omega, \forall t \in \mathcal{T}
\end{align}
where $\alpha_{t,\omega}$ and  $\boldsymbol{\beta}_{t,\omega}$ are defined as :
\begin{align}\label{eq:WScut}
&\alpha_{t,\omega}=p_\omega(\pi_{t,\omega}^{1,r})^\top e_t+p_\omega(\pi_{t,\omega}^{2,r})^\top \ell_{t,\omega}\hspace{1mm},\hspace{1mm}
\boldsymbol{\beta}_{t,\omega}=p_\omega(\pi_{t,\omega}^{1,r})^\top H_{t,\omega}
\end{align}
\end{lemma}
\begin{IEEEproof} 
See Appendix A.
\end{IEEEproof}}

\section{Experiments}
In this section, we present a comprehensive set of experiments to highlight the operational and maintenance performance of the proposed framework.  
In all of the experiments, our test system is composed of two inter-connected microgrids with bidirectional connections to the main grid. Microgrid models are based on a modified IEEE 14-bus test system, as in \cite{chen2011jump}. We use real-world degradation signals to emulate the degradation processes of DERs. We subject machinery to accelerated life tests from new state to failure, and their degradation signals are collected continuously. The details of this physical setup can be found in \cite{gebraeel2005residual}.  
{\ff We consider weekly maintenance decisions. }The DERs maintenance downtime ratio is set to 1:2, meaning that conducting PM takes half of the time to handle a failure.  Details on DERs specification and scenario generation are provided in the supporting document.

{\ff We present three comparative case studies to evaluate the effectiveness and performance of our method.
\begin{itemize}
    \item \textit{Case study 1}: In this case study, we compare the SD-IOM model's performance with a time-based (periodic) maintenance model. Microgrids within the MMG are fully connected and each has 2 MW storage capacity.
    \item \textit{Case study 2}: We analyze the impact of different storage capacities on the fully connected MMG's operations and maintenance schedules. More specifically, we consider a case where microgrids can exchange power with the grid as well as other neighboring microgrids and increase the storage capacity from 0 MW (no-storage) to 4 MW.
    \item \textit{Case study 3}: This case study evaluates the effectiveness of SD-IOM in improving MMG resilience performance in sudden uncertain disruptions. We consider two scenarios where the disruption leads to: 1) disconnection of microgrids from the grid (locally-connected mode), 2) disconnection of microgrids form each other and the grid (islanded-mode).
\end{itemize}}

\subsection{Experimental Framework and Convergence Analysis}
We develop an extensive experimental framework to evaluate the performance of SD-IOM. 
The framework is composed of two main phases: the optimization phase and the evaluation phase. In the optimization phase, we solve SD-IOM  to determine weekly maintenance and hourly operational decisions for a one-year planning horizon. We note that SD-IOM is capable of identifying the DERs' critical condition through dynamic maintenance costs. In the evaluation phase, we assess the performance of SD-IOM against the actual DERs degradation processes. We fix the {\ff optimized} maintenance schedules for the first eight weeks (freezing period). We then simulate the sequence of events that happen following the given maintenance schedules. 

{\ff DERs may experience three outcomes during the planning horizon: (i) \textit{{Unexpected failure}}: The evaluation phase identifies failed DERs by checking whether their corresponding degradation signals reach the failure threshold before the scheduled time of maintenance. These DERs experience outage due to unexpected failure. (ii) \textit{{Planned maintenance}}: For DERs that have not failed during the planning horizon, we determine if maintenance has been scheduled within the planning horizon. If scheduled for maintenance, DERs experience an outage due to planned maintenance. Upon completion of the outage, DERs in categories (i) and (ii) get new degradation signals to characterize the degradation of the replaced component. (iii) \textit{{Uninterrupted degradation}}: Remaining DERs continue degrading without any outage. For these generators, the new observations of the sensor data enable us to update the posterior distribution of the degradation parameters, which, in turn, enables the revision of the degradation signal. We note that the degradation signal is updated for every DER in the system either due to outage or new sensor observation. Finally, the distribution of the remaining life is updated using equation \eqref{eq:R_update}.} 

After updating the DERs availability based on the simulation results, we solve the operational problem considering updated DERs availability as well as revealed renewable generations, loads, market prices, and microgrids' operational modes. This enables us to evaluate the operational performance each week under real MMG conditions. We keep track of reliability, cost, and resilience metrics for each week during the evaluation phase. For example, the SD-IOM model may schedule the DER's maintenance earlier to decrease the failure risks. We refer to the time difference between the maintenance and the failure time as DER's unused life. 

After the evaluation phase, {\ff dynamic maintenance costs of operational DERs are updated using the most recent sensor readings }and the planning horizon is shifted forward to plan the next yearly schedules. {\ff For maintained DERs, new degradation signals from the database are chosen to represent their degradation process after maintenance.} The operating environment in the next optimization phase is then based on what happens during the evaluation phase. We continue this procedure to cover a period of one and a half years. We repeat this procedure by using DERs with different initial ages to ensure a decent comparison. Any metrics presented here is the average of these replicated experiments. 

We compare the cost, reliability, and resilience metrics of SD-IOM with the industry-standard \textit{periodic maintenance} model that enforces a fixed maintenance window based on the age of the DERs. We impose two modifications in the periodic maintenance model: i) setting dynamic maintenance cost functions to zero, ii) enforcing maintenance when the DERs age is between 48 and 52 weeks, which is the optimal window for conducting maintenance given the reliability information. 
 \begin{table}[!htbp]
      \caption{Instance Specifications}
 \centering
\begin{tabular}{p{0.8cm}llccc}
\toprule
&{}&\multicolumn{3}{c}{\textbf{Number of MGs within MMG}} \\
\hline
\textbf{Stage}&\textbf{Number of}& \textbf{2 MGs} &\textbf{3 MGs} &\textbf{4 MGs} \\
\hline
\textbf{First}& \textbf{Binary Vars}    & 400 &   650&1,200  \\
          & \textbf{Constraints}& 352 &  554 &1,059   \\
          \hline
\textbf{Second}& \textbf{Binary Vars }    & 2,721,600 &  4,233,600 &5,745,600  \\
      & \textbf{Continuous Vars}     & 3,024,500
 &  4,687,900 &6,351,000  \\
 &\textbf{Constraints}    & 9,270,000   &  14,774,400 &20,430,000 \\
\bottomrule
\end{tabular}
    \label{table:Nvars}
\end{table}

{\ff We implement the algorithms introduced in Section \ref{section:reformulation} to solve SD-IOM and the periodic model. All experiments are implemented on an Intel Core-i7 2.6 GHz computer using GUROBI 9.0.1. The optimality gap is considered $10^{-3}$ times the absolute value of the lower bound. We investigate the algorithms' convergence performance, both the multi-cut per week and multi-cut per week and scenario versions, on three instances with 2, 3, and 4 interconnected MGs. Table \ref{table:Nvars} illustrates the size of these instances in terms of the number of variables and constraints. Table \ref{table:ALG_0} assesses the algorithm's computational efficiency through the running time in seconds, the number of major iterations, and the number of cuts. We note that the MIP solver cannot optimally solve any instances within two hours time limit. The results show that the introduced decomposition algorithms can significantly decrease the running time in solving SD-IOM problem. Also, results demonstrate the superiority of the multi-cut per week approach in terms of running time and number of cuts.
Table \ref{table:ALG} presents the maximum time incurred for cut generation per iteration in the instances. The results illustrate that as the size of instances increases, the per-week per-scenario increasingly outperforms the per-week method in terms of cut generation time. So, the per-week per-scenario decomposition method offers the use of parallelization methods to further decrease the computational effort specially in large scale instances.}

  \begin{table}[H] \centering
	\caption{Decomposition Algorithms Computational Performance}
	{\begin{tabular}{l  c  c}
		\toprule
		\multicolumn{3}{c}{\textbf{2 MGs}} \\
		\hline
		\textbf{} & \textbf{Per-W} & \textbf{Per-W Per-S}  \\
		\hline
	 \textbf{Running time (s)} & 1121& 1109 \\
	 \textbf{\# Iterations}  & 6& 6\\
	 \textbf{\# Cuts}  & 94& 795 \\
	 		\hline
	 \multicolumn{3}{c}{\textbf{3 MGs}} \\
				\hline
		\textbf{} & \textbf{Per-W} & \textbf{Per-W Per-S}  \\
		\hline
		\textbf{Running time (s)} & 1845 & 2638\\
		\textbf{\# Iterations} & 4& 4\\
		\textbf{\# Cuts}  & 68& 616  \\
		\hline
		
		\multicolumn{3}{c}{\textbf{4 MGs}} \\
				\hline
		\textbf{} & \textbf{Per-W} & \textbf{Per-W Per-S}  \\
		\hline
		\textbf{Running time (s)} & 3007 & 5257\\
		\textbf{\# Iterations} & 5& 5\\
		\textbf{\# Cuts}  & 104& 881  \\
		\bottomrule
	\end{tabular}}
	 \label{table:ALG_0}
\end{table}

 \begin{table}[h] \centering
	\caption{Max Cut Generation Time-Decomposition Algorithms(s) }
{\begin{tabular}{l c c c c c c c} 
\toprule
				   	      \multicolumn{7}{c}{\textbf{2 MGs}} \\
	     \hline
	     {}&\textbf{Iter1}&\textbf{Iter2}&\textbf{Iter3}&\textbf{Iter4}&\textbf{Iter5}&\textbf{Iter6}\\
	     \hline
		\multicolumn{1}{l}{\textbf{Per-W}}& {3.72} & {1.12}&{0.92}& {1} & {0.91}& {1.1}  \\
				\multicolumn{1}{l}{\textbf{Per-W Per-S}}& {1} & {0.52}&{0.5}& {0.4} & {0.56}& {0.45}  \\
				\hline
				   	      \multicolumn{7}{c}{\textbf{3 MGs}} \\
	     \hline
	     {}&\textbf{Iter1}&\textbf{Iter2}&\textbf{Iter3}&\textbf{Iter4}&\textbf{Iter5}\\
	     \hline
		\multicolumn{1}{l}{\textbf{Per-W}}& {7.22} & {2.28}&{1.53}& {0.75} & {-}& {-}  \\
				\multicolumn{1}{l}{\textbf{Per-W Per-S}}& {2.43} & {1.15}&{0.97}& {1} & {-}& {-}   \\
				\hline
				   	      \multicolumn{7}{c}{\textbf{4 MGs}} \\
	     \hline
	     {}&\textbf{Iter1}&\textbf{Iter2}&\textbf{Iter3}&\textbf{Iter4}&\textbf{Iter5}\\
	     \hline
		\multicolumn{1}{l}{\textbf{Per-W}}& {14.37} & {5.72}&{2.23}& {2.18} & {3.38} & {-} \\
				\multicolumn{1}{l}{\textbf{Per-W Per-S}}& {6.58} & {1.21}&{1.84}& {1.1} & {2.87}& {-}   \\
		\bottomrule
		\vspace{-4mm}
\end{tabular}}\label{table:ALG}
\end{table}

\subsection{Experimental Results and Discussions}
In this section, we present the result of our experiments to demonstrate the effectiveness of our approach. All metrics presented in the tables refer to the entire MMG. 
\begin{table}[h] \centering
	\caption{Reliability and cost metrics of MMG}
	{\begin{tabular}{l  c  c} 
		\toprule
		\textbf{Metrics } & \textbf{ Periodic } & \textbf{ SD-IOM }  \\
		\hline
	 \textbf{Exported Power} & 9.23\%& 9.82\% \\
	 \textbf{Imported Power}  & 57.65\%& 56.41\% \\
	 \textbf{Exchanged Power}  & 12.81\%& 11.85\% \\
		\textbf{Curtailed NCL} & 0.03\% & 0.00\% \\
		\textbf{Curtailed CL} & 0.00\% & 0.00\%\\
		\textbf{Curtailed WTs Power}  & 7.98\%& 1.4\%  \\
		\textbf{Curtailed PVs Power} & 7.65\% & 1.76\% \\
		\hline
		\hline
			 \textbf{\# Preventive}  & 22.5& 23  \\
		\textbf{\# Corrective} & 12 & 1.75 \\
	 \textbf{\# Total Outages}  & 34.5& 24.75 \\
		 \textbf{\# Crew Visits} & 21& 17 \\
	 \textbf{Unused Life (wks)}  & 58.15& 16.51 \\
	 \hline
	 \hline
		\textbf{Maintenance Cost} & \$223,950 & \$97,900 \\
		\textbf{Operational Cost} & \$4.352 \scriptsize\textbf{M} & \$4.245 \scriptsize\textbf{M}\\
		\textbf{Total Cost} & \$4.572 \scriptsize\textbf{M} & \$4.339 \scriptsize\textbf{M}\\
		\bottomrule
	\end{tabular}}\label{table:BCase}
\end{table} 
\subsubsection{{\ff Case Study 1}}
The first case compares the performance of SD-IOM with the benchmark (periodic) model. Table \ref{table:BCase} provides the associated reliability and cost metrics. We note that SD-IOM provides significant benefits in terms of costs and reliability. SD-IOM total cost is 5.09\% better than the periodic model, which is a result of significant savings in maintenance and operational costs (56.28\% and 2.45\%, respectively).
Both methods schedule almost the same number of PMs. However, SD-IOM benefits from the sensor-driven predictive model to learn more about the ongoing degradation of DERs and perform maintenance when needed. Consequently, SD-IOM reduces the number of failure instances by 85.41\% compared to the periodic model. Therefore, more generation capacity is available at any time, which leads to a reduction in the operational cost of the SD-IOM model. Comparing the DERs unused life, we see that SD-IOM provides noticeable improvements, i.e., unused life in SD-IOM is 28.39\% of the periodic model. In comparison to the periodic model, a fewer number of DERs outages (due to maintenances) in SD-IOM decreases the need for frequent crew visits by 19\%. 
 \begin{table*}[!htbp] 
	\caption{Average MMG reliability and cost metrics under different storage capacity}
	\centering
	\begin{tabular}{|p{1.1cm}|c| p{1.5cm} p{1.5cm} p{1.2cm} p{1.4cm} p{1.4cm} p{1.4cm} p{1.5cm} p{1.2cm}|} 
		\hline
		\textbf{Storage Capacity} & \textbf{Method} & \textbf{\#Preventive} & \textbf{\#Correctives} & \textbf{\#Total Outages}  &\textbf{\#Crew } &\textbf{Unused Life \scriptsize{(wks)} } & \textbf{Maintenance Cost} &\textbf{Operations Cost} &\textbf{Total Cost} \\
		\hline
		\multirow{2}{*}{\textbf{0 \scriptsize{MW}}} &  \textbf{Periodic}  & 22.5
 & 12
 & 34.75 &21 &58.16 & \$223,950 & \$4.447 \scriptsize\textbf{M} & \$4.666 \scriptsize\textbf{M}  \\
		&\textbf{SD-IOM} & 23.5 & 1.5 & 25 &17.75&16.52 & \$100,550 & \$4.303 \scriptsize\textbf{M} & \$4.400 \scriptsize\textbf{M}   \\
		\hdashline
		\multirow{2}{*}{\textbf{2 \scriptsize{MW}}} &  \textbf{Periodic}  & 22.5
 & 12 & 34.5& 21 &58.15 & \$223,950 & \$4.352 \scriptsize\textbf{M} &  \$4.572 \scriptsize\textbf{M} \\
		& \textbf{SD-IOM} & 23
 & 1.75 & 24.75& 17& 16.51 & \$97,900 & \$4.245 \scriptsize\textbf{M}& \$4.399 \scriptsize\textbf{M} \\
		\hdashline
		\multirow{2}{*}{\textbf{4 \scriptsize{MW}}} &  \textbf{Periodic}  & 22.5
 & 12 & 34.5 & 20.75& 58.12 &\$223,900& \$4.318 \scriptsize\textbf{M}& \$4.538 \scriptsize\textbf{M}\\
		& \textbf{SD-IOM} & 22.75 & 1.75
 &24.5&17&15.88 & \$95,900 & \$4.222 \scriptsize\textbf{M}& \$4.314 \scriptsize\textbf{M} \\
		\hline
		\hline
		\textbf{Storage Capacity} & \textbf{Method} & \textbf{Curtailed WTs Power} & \textbf{Curtailed PVs Power} & \textbf{Imported Power} &\textbf{Exported Power} & \textbf{Exchanged Power} &\textbf{Curtailed NCL} &\textbf{Curtailed CL} &\textbf{Curtailed L Cost}\\
		\hline
		\multirow{2}{*}{\textbf{0 \scriptsize{MW}}} &  \textbf{Periodic}  & 9.53\%& 7.77\%& 59.06\% & 5.38\% & 13.65\% & 0.12\%&0.00\% &\$42,787.5 \\
		&\textbf{SD-IOM}  & 1.40\%& 1.77\%& 55.93\%& 7.52\% & 11.52\% & 0.04\%&0.00\%&\$14,025\\
		\hdashline
		\multirow{2}{*}{\textbf{2 \scriptsize{MW}}} &  \textbf{Periodic}  & 7.98\%
 & 7.65\% & 58.97\% & 7.9\% & 14.16\%&0.03\%&0.00\% &\$9,742.5 \\
		& \textbf{SD-IOM} & 1.40\%& 1.76\%& 56.57\% & 9.89\% & 12.25\% & 0.00\%&0.00\%&\$450.00\\
		\hdashline
		\multirow{2}{*}{\textbf{4 \scriptsize{MW}}} &  \textbf{Periodic}  & 7.89\%
 & 7.65\% & 59.14\% & 9.64\% & 13.84\%&0.00\%&0.00\% &\$536.25 \\
		& \textbf{SD-IOM} & 1.40\%
 & 1.77\% & 56.93\% & 11.47\% & 11.49\% & 0.00\%&0.00\%&\$0.00\\
		\hline		
	\end{tabular}
	\label{table:SD}
\end{table*} 
Reduction in the number of failure instances, as well as unused life of DERs in SD-IOM, results in significant operational advantages. More available renewable DERs in SD-IOM, i.e., WTs, and PVs, means lower renewable curtailment. Specifically, periodic model curtails 7.98\%, and 7.65\% of wind and solar power, respectively, due to outages while SD-IOM curtailments are only  1.4\% and 1.76\%. In terms of MMG power interaction, in SD-IOM microgrids, in total, utilize 11.85\% of transmission lines capacities to exchange power with each other while in the periodic model, this amount increases by 0.96\%. Lower imported power from the main grid,  by 1.24\%, and higher exported power to the main grid along with lower exchanged power within the MMG show higher autonomy of individual microgrids and MMG in general. Higher available capacities provide more support from the MMG for the main grid and lower the microgrids' dependency on each other and the grid alike. 
\subsubsection{{\ff Case Study 2}}
In this section, we analyze the SD-IOM and periodic model performance under different storage capacities. We alter the storage capacity from 0 MW (no storage) to 4 MW to study the impact of maintenance policies on the reliability \& cost metrics. Table \ref{table:SD} shows that both models are capable of reducing the operational, maintenance, and total 
costs as the storage capacity increases.
However, even with 4 MW storage capacity, the periodic model's operational, maintenance, and total costs cannot compete with the SD-IOM in the no storage case. 
In response to higher storage capacity, SD-IOM tries to provide more power generation available. Consequently, the model reduces DERs outages by decreasing the number of PMs. This leads to slightly higher failure instances but lowers the unused life of DERs by 3.87\% to raise power generation at hand. Likewise, the periodic model reacts to higher storage capacity by reducing the number of outages. However, without access to DERs health condition, it cannot deviate too much from the recommended maintenance windows. As a result, failure instances remain the same, while the decline in unused DERs life is only 0.06\%.
Moreover, the frequency of maintenance crew visits decreases with the reduction in the number of outages. By reducing the unused life of DERs, the renewable curtailment decreases in the periodic model. Nevertheless, the SD-IOM model performance is still superior to the periodic model, by curtailing at most 1.4\% of renewable generation. Both models take advantage of higher storage capacity to perform energy arbitrage, power balancing, and ancillary services. In the SD-IOM model, as the storage capacity increases, the imported power from the main grid and the exported power to the main grid increases by 1.78\% and 52.52\%.
With the presumed MMG setting, both models are capable of satisfying the critical loads under normal operational mode. We highlight that the periodic model with 2 MW storage manages to obtain what the SD-IOM model with no storage achieves in terms of load curtailment. 
\subsubsection{{\ff Case Study 3}}
This final case study analyzes the value of sensor-driven maintenance on enhancing operational resilience. The resilience performance of MMG is studied under two different scenarios: i) locally-connected mode ii) islanded mode. In the first scenario, we consider a case that a sudden uncertain disruption leads to MMG disconnection from the main grid, i.e., the transferred power from the main grid to microgrids is zero. In the second scenario, each microgrid has no power transactions with other entities. 
We assume that microgrids' components do not expose to disruption. Many metrics have been proposed in the literature to assess resilience. {\ff Here we define resilience as the ability of MMG to maintain its performance quality during the disruption. Let us define $q^*(t)$ as the as-planned operational factor of MMG and {\ff$\tilde{q}^*(t)$} as the operational factor during the recovery period. Then the resilience of MMG at time $t$ is equal to $\Psi(t)=q^*(t)/\tilde{q}^*(t)$. The resilience ranges from 0\% up to 100\% , where 100\% means no degradation in the quality of operational factor. We consider MMG's capability in satisfying critical and non-critical loads as well as maintaining the operational profitability as three important operational factors of MMG. We evaluate the expected degradation in the performance quality of MMG by the expected resilience loss (ERL).} The ERL measure is modified from the traditional resilience loss\cite{hosseini2016review}. This measure compares the operational performance of disrupted MMG to the as-planned operational performance. The ERL measure is defined as follows:\vspace{-3mm}
\begin{table*}[ht]
\caption{Expected resilience loss of MMG during disruption}
\centering
\begin{tabular}{l|  cc cc | cc cc}
\toprule
 \textbf{Storage Capacity}& \multicolumn{4}{|c|}{2 MW} &\multicolumn{4}{c}{0 MW}\\
\midrule
& \multicolumn{2}{|c}{\quad MMG Locally-connected \quad} &\multicolumn{2}{c|}{MGs Islanded} & \multicolumn{2}{c}{\quad MMG Locally-connected \quad} &\multicolumn{2}{c}{MGs Islanded}\\
\cmidrule(lr){2-3} \cmidrule(lr){4-5} \cmidrule(lr){6-7} \cmidrule(lr){8-9} 
     &\scriptsize{Periodic}   & \scriptsize{SD-IOM}   &\scriptsize{Periodic}   & \scriptsize{(SD-IOM)} &\scriptsize{Periodic}   & \scriptsize{(SD-IOM)} &\scriptsize{Periodic}   & \scriptsize{SD-IOM}\\
        \midrule
		\textbf{Critical Loads} & 0.13\%& 0.01\%& 0.68\% & 0.22\%& 0.18\%  & 0.04\%& 0.89\%&0.42\% \\
		\textbf{Non-Critical Loads} &6.33\% &1.89\% &13.78\%&7.10\%& 8.38\%  & 3.53\%& 13.93\% & 7.69\% \\
		\textbf{Operational Costs} & 29.12\% & 15.63\%&54.41\%&32.25\%& 38.15\%  & 22.19\%& 58.29\% & 37.66\% \\
\bottomrule
\end{tabular}\label{table:resilience}
\label{table:RL}
\end{table*}
\begin{figure}[t!]
    \centering 
    \includegraphics[width=0.5\textwidth]{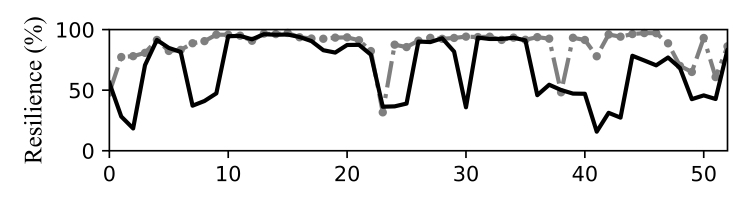}
    \includegraphics[width=0.5\textwidth]{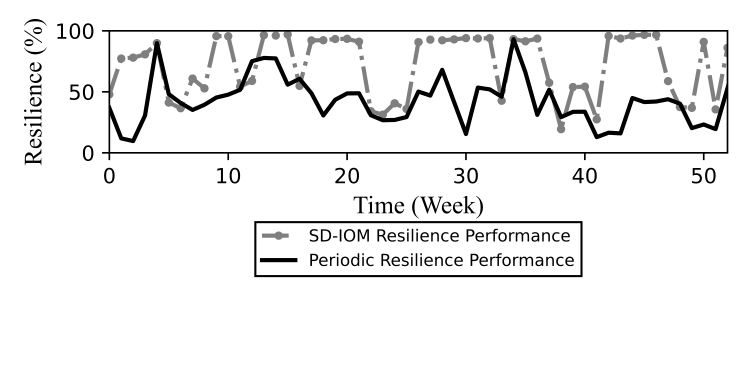}  
    \vspace{-15mm}
             \caption{MMG resilience performance in the (top) locally-connected  and (bottom) islanded Mode -2MW storage capacity} 
                     \label{fig:resilience} 
\end{figure}
{\ff
\begin{equation}\label{eq:ResMetric}
ERL=\frac{\sum\limits_{t=1}^{T} p_{t}\sum\limits_{t_0=t}^{t+t^d}\big[1-\Psi(t_0)\big]}{t^d}=\frac{\sum\limits_{t=1}^{T} p_{t}\sum\limits_{t_0=t}^{t+t^d}\big[1-\frac{q^*(t_0)}{\tilde{q}^*(t_0)}\big]}{t^d}
\end{equation}}

where $t_0$ is the disruption start time, $t^d$ is the disruption duration and $p_t$ is the probability of disruptive event at period $t$.
ERL metric indicates the expected decline in the performance quality of MMG due to disruptions {\ff over a period $T$. Lower values of ERL indicate higher resilience performance.}

To evaluate the average resilience performance of MMG, we assume that the disruption can happen at the beginning of any week, and the damage would be restored by the end of the week. Table \ref{table:RL} shows the ERL {\ff in the operational cost as well as satisfying critical and non-critical loads.  We compare the ERLs under two different settings in which microgrids within MMG has no storage and 2 MW storage capacity. The result shows the proposed framework capability in  achieving the lowest expected resilience loss in all cases. In particular, our model provides much lower ERL in terms of satisfying loads (both critical and non-critical). In the locally-connected mode, we observe that the ERLs of SD-IOM in satisfying critical and non-critical loads are $77.7\%$ and $57.9\%$ lower than the corresponding values in the benchmark model. Correspondingly, the SD-IOM's ERL in operational cost is $41.83\%$ lower than the benchmark model in the locally-connected mode. Both models experience higher ERLs in the microgrids sudden islanded mode, but SD-IOM still outperforms the periodic model. Introducing more storage capacity decreases the ERLs in both models. However,} in both locally-connected and islanded mode, the ERLs of the periodic model with 2 MW storage are still higher than the corresponding ERLs of the SD-IOM model with no storage. In total, in the presence of sensor information, microgrids within the MMG have a higher capability to support each other and work independently, if necessary.

Figure \ref{fig:resilience} provides the resilience performance of models in terms of operational costs under locally-connected and islanded mode. Operational resilience during disruption is defined as the operational cost in the normal mode divided by the operational cost in the emergency mode. Note that the SD-IOM model, in general, has better resilience performance in both locally-connected and islanded mode.
A sudden islanded mode leads to a sharp decline in the performance of the periodic model, while SD-IOM model, in some periods, is still capable of maintaining resilience around one.  
\section{Conclusion}
In this paper, we propose an integrated sensor-driven framework to improve MMG operations and maintenance. We address unique challenges associated with modeling and computation for deploying condition-based maintenance in an MMG setting. The proposed approach provides significant benefits in terms of asset reliability and operational metrics, such as cost, renewable penetration, and resilience.
\bibliographystyle{ieeetr} 
\bibliography{main} 
\appendices
{\ff\section{Proof of Lemma 1}
The objective of this section is to prove the validity of the optimality cuts in per week, and per-week per-scenario variants of SD-IOM. We denote $\boldsymbol{z}^r$ as an optimal solution of the first stage problem at iteration $r$. This solution is used as a fixed parameter in the sub-problems. We present the cut generation procedure for the two variants of SD-IOM as follows:
\begin{itemize}
    \item 
\textit{Per Week Optimality Cuts:}

We solve the weekly decomposed relaxed sub-problem for each week $t\in\mathcal{T}$:
\begin{subequations}
\begin{align}
 \mathcal{R}_t(\boldsymbol{z}_t^r)=\underset{\boldsymbol{x}_t, \boldsymbol{y}_t}{\text{min}} \hspace{1mm} \sum\limits_{\omega\in\Omega}p_\omega(\boldsymbol{q}_t^\top \boldsymbol{x}_{t,\omega}+\boldsymbol{b}_{t,\omega}^\top \boldsymbol{y}_{t,\omega})\hspace{2cm}\nonumber\\ 
\text{s.t.}\hspace{1mm} \boldsymbol{E}_t\boldsymbol{x}_{t,\omega} +\boldsymbol{D}_t\boldsymbol{y}_{t,\omega}\leq \boldsymbol{e }_t-\boldsymbol{H}_{t,\omega}\boldsymbol{z}^r_t\nonumber\hspace{1mm},\hspace{1mm}\forall\omega\in\Omega\\ 
\boldsymbol{L}_t\boldsymbol{x}_{t,\omega}+\boldsymbol{G}_t\boldsymbol{y}_{t,\omega}\leq \boldsymbol{\ell}_{t,\omega}\hspace{1.39cm},\hspace{1mm}\forall \omega \in \Omega\nonumber
\end{align}
\vspace{-0.7cm}
\begin{align}
\hspace{1.8cm}\boldsymbol{x}_{t,\omega} \in [ 0,1 ]^{\big(3 G^{\mathrm{nr}}+2B+2M+\sum\limits_{m=1}^{M}2N(m)\big) \cdot H\cdot|\Omega|}\nonumber\\\boldsymbol{y}_{t,\omega} \geq 0,\forall \omega \in \Omega\nonumber\hspace{2.6cm}
\end{align}
\end{subequations}
We represent the dual multipliers associated with the optimal solution of the sub-problem at iteration $r$ with $\bm{\pi}^{i,r}_{t,\omega}\hspace{1mm},\hspace{1mm}i=1,2$. We define:
\begin{align}
&\alpha_t=\sum\limits_{\omega\in\Omega}\big((\pi_{t,\omega}^{1,r})^\top e_t+(\pi_{t,\omega}^{2,r})^\top \ell_{t,\omega}\big)\hspace{1mm},\hspace{1mm}\forall t\in\mathcal{T}\nonumber\\
&\boldsymbol{\beta}_t=\sum\limits_{\omega\in\Omega}(\pi_{t,\omega}^{1,r})^\top H_{t,\omega}\hspace{1mm},\hspace{1mm}\forall t\in\mathcal{T}\nonumber\end{align}
Per week optimality cuts can be generated as follows:
\begin{align}
&\eta_{t}\geq \alpha_{t}- \boldsymbol{\beta}_{t}\boldsymbol{z}_t\hspace{1mm},\hspace{2mm}\forall t \in \mathcal{T}\nonumber
\end{align}
where $\eta_{t}$ is a free variable. 
\item \textit{Per Week \& Scenario Optimality Cuts:}

In this method, we decompose the relaxed sub-problem per week and per scenario and then solve the following sub-problem:
\begin{subequations}
\begin{align}
\mathcal{R}_{t,\omega}(\boldsymbol{ z }_t^r)=\underset{\boldsymbol{x}_t, \boldsymbol{y}_t}{\text{min}} \hspace{3mm} \boldsymbol{q}_t^\top \boldsymbol{x}_{t,\omega}+\boldsymbol{b}_{t,\omega}^\top \boldsymbol{y}_{t,\omega}\hspace{1.4cm}\nonumber\\ 
\hspace{.4cm}\text{s.t.}\hspace{2mm} \boldsymbol{E}_t\boldsymbol{x}_{t,\omega} +\boldsymbol{D}_t\boldsymbol{y}_{t,\omega}\leq \boldsymbol{e }_t-\boldsymbol{H}_{t,\omega}\boldsymbol{z}^r_t\nonumber\hspace{.20cm}\\ 
\boldsymbol{L}_t\boldsymbol{x}_{t,\omega}+\boldsymbol{G}_t\boldsymbol{y}_{t,\omega}\leq \boldsymbol{\ell}_{t,\omega}\hspace{1cm} \nonumber
\end{align}
\vspace{-0.7cm}
\begin{align}
\boldsymbol{x}_{t,\omega} \in [ 0,1 ]^{\big(3 G^{\mathrm{nr}}+2B+2M+\sum\limits_{m=1}^{M}2N(m)\big) \cdot H},\boldsymbol{y}_{t,\omega} \geq 0\nonumber
\end{align}
\end{subequations}
Let $\bm{\pi}^{i,r}_{t,\omega}$ and $\bm{\pi}^{2,r}_{t,\omega}$ denote the dual multipliers associated with the first and second constraints, respectively. We define:
\begin{align}
&\alpha_{t,\omega}=p_\omega(\pi_{t,\omega}^{1,r})^\top e_t+p_\omega(\pi_{t,\omega}^{2,r})^\top \ell_{t,\omega}\hspace{1mm},\hspace{1mm}\forall t\in\mathcal{T},\forall\omega\in\Omega\nonumber\\
&\boldsymbol{\beta}_{t,\omega}=p_\omega(\pi_{t,\omega}^{1,r})^\top H_{t,\omega}\hspace{2.2cm},\hspace{1mm}\forall t\in\mathcal{T},\forall\omega\in\Omega\nonumber
\end{align}
Per week optimality cuts can be generated as follows:
\begin{align}
&\eta_{t,\omega}\geq \alpha_{t,\omega}- \boldsymbol{\beta}_{t,\omega}\boldsymbol{z}_t\hspace{1mm},\hspace{2mm}\forall \omega \in \Omega, \forall t \in \mathcal{T}\nonumber
\end{align}
\end{itemize}
{\ff We note that there is no need to add feasibility cuts since sub-problems are always feasible for any solution $\boldsymbol{z}_t$ from the master problem (i.e. complete recourse).} 
}
\end{document}